\documentclass[aps,prl,twocolumn]{revtex4-2}
\usepackage{epsfig,amssymb,amsmath,amsthm,amsfonts,amsbsy,mathrsfs}
\usepackage{graphicx}
\usepackage{color}
\usepackage{xcolor,hyperref}
\hypersetup{
   colorlinks,
   linkcolor={blue!50!black},
   citecolor={blue!50!black},
   urlcolor={blue!80!black}
}

\usepackage{tikz}
\usetikzlibrary{arrows,shapes,chains}

\definecolor{cream}{RGB}{222,217,201}

\begin{document}
\title{Local Plastic Response and Slow Heterogeneous Dynamics of Supercooled Liquids}
\author{Yan-Wei Li}
\email{yanweili@bit.edu.cn}
\affiliation{Key Laboratory of Advanced Optoelectronic Quantum Architecture and Measurement (MOE), School of Physics, Beijing Institute of Technology, Beijing, 100081, China}
\author{Yugui Yao}
\email{ygyao@bit.edu.cn}
\affiliation{Key Laboratory of Advanced Optoelectronic Quantum Architecture and Measurement (MOE), School of Physics, Beijing Institute of Technology, Beijing, 100081, China}
\author{Massimo Pica Ciamarra}
\email{massimo@ntu.edu.sg}
\affiliation{Division of Physics and Applied Physics, School of Physical and
Mathematical Sciences, Nanyang Technological University, Singapore 637371, Singapore}
\affiliation{
CNR--SPIN, Dipartimento di Scienze Fisiche,
Universit\`a di Napoli Federico II, I-80126, Napoli, Italy
}
\affiliation{
CNRS@CREATE LTD, 1 Create Way, \#08-01 CREATE Tower, Singapore 138602
}
\date{\today}

\begin{abstract}
We demonstrate, via numerical simulations, that the relaxation dynamics of supercooled liquids correlates well with a plastic length scale measuring a particle's response to impulsive localized perturbations and weakly to measures of local elasticity.
We find that the particle averaged plastic length scale vanishes linearly in temperature and controls the super-Arrhenius temperature dependence of the relaxation time.
Furthermore, we show that the plastic length scale of individual particles correlates with their typical displacement at the relaxation time. In contrast, the local elastic response only correlates with the dynamics on the vibrational time scale.
\end{abstract}
\maketitle
When supercooled, liquids develop a slow relaxation dynamics~\cite{EdigerReview, StillingerNature, BerthierReview} that proceeds through a sequence of activated events~\cite{Ciamarra2015SM}.
These relaxation events occur in preferential locations fixed by the particle configuration~\cite{Widmer-Cooper2004}, rather than randomly in space. 
Furthermore, these events cluster in space and propagate through a facilitation mechanism~\cite{Candelier2010, Keys2011, Pastore2015c}, making the dynamics spatially heterogeneous on the relaxation time scale~\cite{EdigerReview, StillingerNature, BerthierReview, Harrowell_NP, ChandlerPRX, tong_Tanaka, Hu_wang}.
It is currently unclear what features of a particle configuration identify the regions where relaxation events occur with higher probability.

Previous works suggested that the local structure surrounding each particle determines its possible involvement in irreversible relaxation events and analysed this structure introducing physically motivated structural parameter~\cite{Tanaka2008PRL,patrick2008, Patrick2014, Akira2018, Li2016, Tanaka2008PRL, Tanaka2011_natureMaterial} or via machine learning approaches~\cite{AJLiu_PRL2015, AJLiu_NP2016}.
These structural parameters are system dependent as affected, e.g., by the particles' shape.
Alternatively, local elastic rather than geometric properties may have predictive ability, if the established correlation~\cite{Dyre2006, Larini2008, PazminoBetancourt2015} between macroscopic elastic properties and relaxation dynamics also holds at the microscopic scale.
Considered measures of local elasticity include the Debye-Waller factor (DWF), its harmonic approximation, and parameters probing particles' vibrational motion~\cite{Buchenau1992a, Harrowell_NP, softModes_polymer, SoftModes_Nonaffine, HarrowellPRL2006}. 
However, the correlation of a particle's local elastic properties with its involvement in local relaxation processes is debated.
Indeed, a particle's vibrational amplitude correlates highly with its short-time displacement~\cite{Harrowell_NP} but weakly with its displacement evaluated at the relaxation time scale~\cite{Patrick2014}.
Besides, correlations between local elastic vibrations and plastic relaxation events entail non-obvious assumptions~\cite{Kumar2021} on the shape of the free energy basin the system is transiently confined.


The response of individual particles to externally applied forces is an alternative approach to characterize local mechanical properties.
The response to small forces, which probes the linear elastic regime, revealed the spatial heterogeneity of the local elastic moduli in metallic glasses~\cite{NM2011Elastic} and dense colloidal suspensions~\cite{Weeks_linearElastic}.
Larger forces induce plastic flow and allow for microrheology investigations~\cite{Habdas2004, Puertas2014, Furst2018, Senbil}.
Recently, the response to transient applied forces has been used to probe the emergence of caging in colloidal suspensions~\cite{GranickNature2020} and energy absorption in soft colloids~\cite{Jasper_SA,Lucio_PNAS}.

In this Letter, we demonstrate that the plastic response induced by transient localized perturbations defines a length, $\xi$, that correlates with the relaxation dynamics at the macroscopic and at the individual particle level.
The average plastic length vanishes linearly with the temperature and regulates the super-Arrhenius divergence of the relaxation time, $\tau = \tau_0 \exp(\xi_0/\xi)$. A particle's local plastic response correlates with its typical displacement at the relaxation time. 
On the contrary, the local elastic response only correlates with a particle's displacement at shorter times, on the vibrational time scale.
Our findings demonstrate that structural relaxation proceeds via local rearrangements that are weakly related to the local elastic response but well connected to the local plastic response induced by transient perturbations.

We consider the Kob-Andersen~\cite{KA_94, mKA}, 65(A):35(B) binary Lennard-Jones mixture of $N=2000$ particles in two spatial dimensions. 
The interaction potential is $U_{\alpha\beta}(r) =4\epsilon_{\alpha\beta}[(\frac{\sigma_{\alpha\beta}}{r})^{12} - (\frac{\sigma_{\alpha\beta}}{r})^{6} + C_{\alpha\beta}]$ for $r \leq r_{\alpha\beta}^{c} = 2.5\sigma_{\alpha\beta}$ and $U_{\alpha\beta}(r) = 0$ otherwise, $\alpha, \beta \in$ $\{\rm {A, B}\}$. 
We set $\sigma_{AB}/\sigma_{AA}=0.8$, $\sigma_{BB}/\sigma_{AA}=0.88$, $\epsilon_{AB}/\epsilon_{AA}=1.5$, $\epsilon_{BB}/\epsilon_{AA}=0.5$, and fix $C_{\alpha \beta}$ so that the potential vanishes continuously at the cutoff. 
Length, energy and time are recorded in units of $\sigma_{AA}$, $\epsilon_{AA}$ and $\sqrt{m\sigma_{AA}^{2}/\epsilon_{AA}}$, respectively. 
We perform Langevin dynamics simulations using {\scshape lammps}~\cite{Lammps}, numerically integrating the equations of motion, $m\mathbf{\ddot{r}}_{i}=-\mathbf{\nabla}_{i} \sum_{j(\neq i)} U_{\alpha \beta}(r_{ij})-\gamma \mathbf{\dot{r}}_{i}+\mathbf{\eta}_{i}(t)$, where $\mathbf{r}_{i}$ is the position of the $i$th particle, $r_{ij}$ is the inter-particle distance, and $\gamma=1$ is the friction coefficient.
$\mathbf{\eta}_{i}$ is a random noise, satisfying $\langle\mathbf{\eta}_{i}(t)\mathbf{\eta}_{j}(t')\rangle=2k_{B}T\gamma\delta_{ij} \delta_{t-t'}\mathbf{1}$, with $\mathbf{1}$ the unit tensor. 
This model has been shown to be a good glass former~\cite{mKA}, the relaxation time developing a super-Arrhenius temperature dependence below the onset temperature $T_{\rm onset}=0.7$~\cite{Yanwei_MassimoPNAS,SM}, for number density $\rho=1.2$.
We recap the main features of the relaxation dynamics of this model in the Supplementary Material (SM)~\cite{SM}.

We probe the local mechanical response by exerting to a  particle $i$ of an equilibrium configuration a force of magnitude $f$ and random orientation in the time interval $0<t< t_{\rm p}$, mimicking recent experimental studies~\cite{GranickNature2020, Weeks_linearElastic, Lucio_PNAS, Jasper_SA}.
The response of the system depends on $f$, $t_p$ and the damping parameter $\gamma$.
In the following, we concentrate on the $f$ dependence of the response at fixed $\gamma = 1$ and $t_p=0.1$ corresponding to $\approx 1/10$ of the vibrational timescale. 
In the SM~\cite{SM}, we show that results are robust with respect to changes in $\gamma$ and $t_p$ as long as the perturbation triggers an irreversible response.

\begin{figure}[tb]
 \centering
 \includegraphics[angle=0,width=0.43\textwidth]{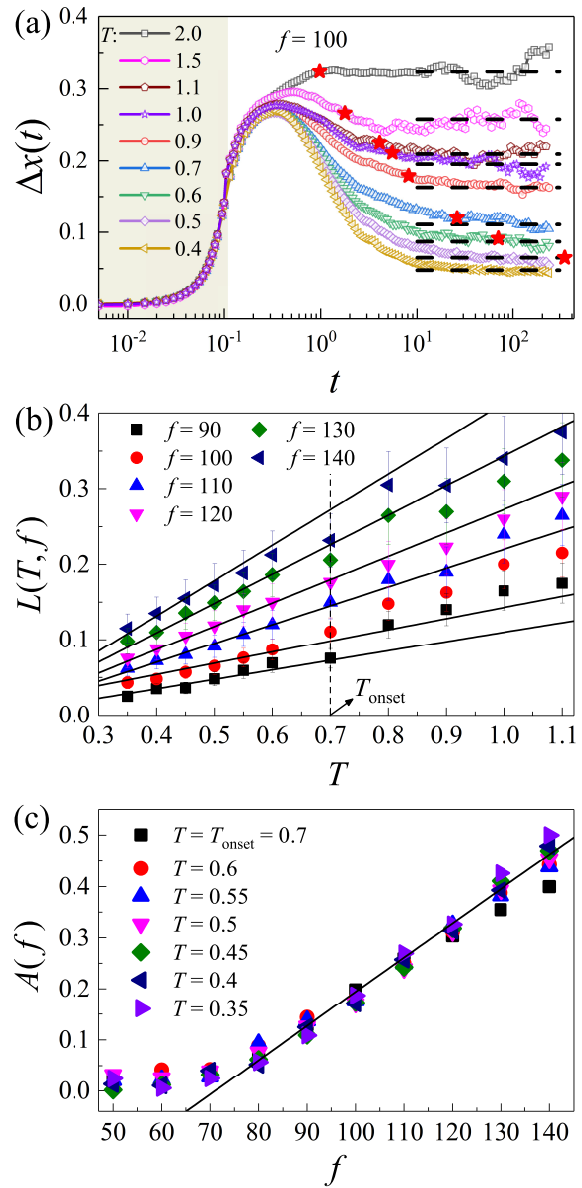}
 \caption{
  (a) Time dependence of the average displacement of a particle in response to a force acting on the time window identified by the shaded region. The dashed lines mark the asymptotic displacement values, $L(T,f)$. The filled red stars mark the $\Delta x(\tau_{\rm CR})$ for $T\geq0.5$.
 (b) The asymptotic displacement grows linearly in temperature below the onset temperature marked by the vertical dashed line, $L \propto (T-T_0)$.
 (c) A plot of $A(f) = L(T,f)/(T-T_0)$ reveals that, below the onset temperature, the asymptotic displacement grows linearly in $f$ for large enough forces. 
 \label{fig:DeltaX}
}
\end{figure}

We illustrate in Fig.~\ref{fig:DeltaX}(a) the time dependence of the displacement $\Delta x(t) = \langle \Delta x_i(t_{\rm obs}) \rangle_{\theta,N_{\rm r},c}$ induced by a local force of magnitude $f$.
The displacement is averaged over $160$ forces differing in their random orientation $\theta$ for each particle, and then further averaged over $N_{\rm r}=200$ randomly selected particles and $c=8$ independent configurations.
The displacement attains a constant value $L$ for a transient which approximately ends when the mean square displacement becomes of order $L^2$.
Below the onset temperature, this transient begins at $t \simeq 2\times 10^2$, corresponding to roughly  $t_{\rm obs} \simeq 200$ vibrational times.

The time dependence of the displacement informs on the elastic properties of the systems.
Indeed, in a viscous medium, the displacement induced by a transient localized force increases monotonically and saturates to a plateau value, $L  \propto f$. 
In Fig.~\ref{fig:DeltaX}(a), where dashed lines mark the value of $L$, we observe this behaviour at the largest of the considered temperatures.
In contrast, in a viscoplastic medium, the displacement reaches a maximum and then saturates to a smaller $L$ value, as we observe at the other considered temperatures. 
In the presence of a purely elastic response, the final displacement $L$ would vanish.

The magnitude of the average force-induced displacement $L(T,f)$ (see Fig. S4~\cite{SM} for the distribution of the individual particle displacements) depends on the applied force and temperature.
Below the onset temperature the asymptotic displacement increases linearly with the temperature, $L(T,f) \propto (T-T_0)$ and with the applied force, $L \propto (f-f_0)$, as illustrated in Fig.~\ref{fig:DeltaX}(b) and (c).
Here, $T_0 \simeq 0.12$ and $f_0$ is a temperature independent threshold value.
This scaling does not hold at higher temperatures (see Fig. S3~\cite{SM}), and indeed, a small deviation already occurs at the onset temperature in Fig.~\ref{fig:DeltaX}(c).
The need of applying a minimum force to induce irreversible particle motion is consistent with previous results~\cite{Senbil}.

To rationalize the force dependence of $L$, we investigate the response to the application of localized forces of systems brought in energy minima, in the SM~\cite{SM}.
This investigation suggests that the temperature independence of $f_0$ reflects the weak dependence of the energy landscape on the temperature of the parent liquid, in the supercooled regime, and that $L$ has a linear dependence on $f$ as it results from the accumulation of $n(f)\propto (f-f_0)$ plastic displacements of typical size $\langle l\rangle $~\cite{SM}.

Summarizing, Fig.~\ref{fig:DeltaX} shows that in the regime of low-temperature and forces large enough to trigger plastic rearrangements, the force-induced displacement varies as $L(T,f) = \xi(T) (1-f/f_0)$, with $\xi(T) \propto (T-T_0)$ a temperature dependent length scale.
Next, we show that this plastic length regulates the relaxation dynamics.

We probe the relaxation dynamics using cage-relative (CR) measures to filter out the influence of long-wavelength fluctuations~\cite{MWPrl, Shiba, Weeks_longwave, Keim_MW, Yanwei_MassimoPNAS} that lead to an underestimation of the structural relaxation time in two spatial dimensions.
While standard relaxation measures focus on the actual displacement of each particle, CR measures focus on the displacement of a particle with respect to the average displacement of its neighbours, which we identify via the Voronoi construction.
This definition makes CR measures unaffected by the collective translations that characterise long-wavelength fluctuations.

\begin{figure}[tb]
 \centering
 \includegraphics[angle=0,width=0.48\textwidth]{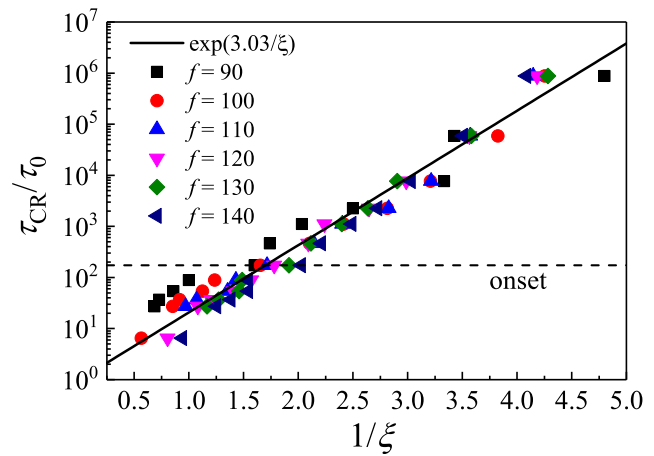}
 \caption{
 The cage-relative relaxation time grows exponentially with the reciprocal plastic length $1/\xi(T) = (1-f/f_0)(1-T/T_0)/L(T,f)$. Data refer to $\rho = 1.2$.
\label{fig:scaling}
}
\end{figure}

We estimate the CR relaxation time, $\tau_{\rm CR}(T)$, which is proportional to the shear viscosity~\cite{Yanwei_MassimoPNAS, Hayato_PRL2019}, from the decay of the CR self-scattering function evaluated at the wave-vector corresponding to the first peak of the structure factor.
Figure~\ref{fig:scaling} illustrates that, below the onset temperature, the CR relaxation time grows exponentially with the inverse plastic length scale $\xi(T)$,
\begin{equation}
    \tau_{\rm CR}(T) = \tau_0 \exp\left(\frac{\xi_0}{\xi(T)}\right) 
    \label{eq:tau}
\end{equation}
with $\tau_0$ a microscopic vibrational timescale and $\xi_0$ a constant.
Given that $\xi \propto (T-T_0)$, Eq.~\ref{eq:tau} corresponds to a Vogel-Fulcher-Tammann (VFT) dependence of the relaxation time on temperature, which we illustrate in Fig. S2 of the SM~\cite{SM}.

We stress that other functional forms may describe the super-Arrhenius temperature dependence of the relaxation time. As such, we do not want to give fundamental significance to the fitting parameters.
Yet, since $\tau = \tau_0 \exp(\Delta F/T)$, where $\Delta F(T)$ is the free energy barrier of the elementary relaxation events, our finding demonstrates a connection between the plastic length scale $\xi(T)$ and $\Delta F(T)$, i.e., $\Delta F \propto 1/\xi$.
This is our first main result.

A VFT-like divergence of the relaxation time is predicted by entropic theories of the glass transition, such as the early Adam-Gibbs~\cite{AG} scenario and the subsequent Random First Order Transition theory~\cite{RFOT}.
According to these theories, the super-Arrhenius dynamics follows from the vanishing of the configurational entropy, or equivalently from the divergence of a static length which measures the size of loosely defined cooperative rearranging regions~\cite{Stevenson_Wolynes_NP,Stringlike_Glotzer_Kob}.
Previous works supported the existence of a growing static length on cooling~\cite{Biroli2008, Tanaka2008PRL, Tanaka2011_natureMaterial, Sastry_review,tong_Tanaka,Ganapathy_NP,Han_sa}.
These theories are consistent with our results only if the dynamical slowdown involves both a decreasing and an increasing length scale.
This scenario may occur if relaxation events have a core-corona structure, with the core and corona size respectively decreasing and increasing on cooling.
In this respect, we remark that particles undergoing large displacements in a relaxation event become increasingly localized~\cite{Coslovich2019,Ji2020} on cooling.

Elastic models of the glass transitions~\cite{Hall1998, Dyre2006} relates $\Delta F$ to a vanishing length scale, the DWF, e.g., $\Delta F \propto 1/\langle u^2 \rangle^{\alpha/2}$ with
$\alpha$ a phenomenological parameter describing cage anisotropy~\cite{Simmons2012, PazminoBetancourt2015} or related functional forms~\cite{Buchenau1992a,Larini2008}. 
While both $\xi$ and $\langle u^2 \rangle$ decrease on cooling, these lengths are conceptually different. $\xi$ relates to the plastic response and it appears to vanish at a finite temperature. 
In contrast, $\langle u^2 \rangle$ probes the elastic, possibly anharmonic, vibrational motion and only vanishes at zero temperature.
Indeed, to recover a super-Arrhenius behaviour, earlier works~\cite{Buchenau1992a} suggested the relaxation time relates to the difference between $\langle u^2 \rangle$ and its expected crystalline value.

We now turn our attention to the spatial and temporal heterogeneity of the relaxation dynamics~\cite{EdigerReview, StillingerNature, BerthierReview},
which we quantify by resorting to the CR mean square displacement in the iso-configurational ensemble~\cite{Widmer-Cooper2004}.
We evaluate $\langle\Delta r^2_{i, \rm CR}(t)\rangle_{\rm iso}$ by averaging the mean square displacement of each particle over $128$ equilibrium simulations sharing the same initial particle configuration and differing in the particles' momenta, we randomly sample from the relevant equilibrium distribution.
Particles with a large $\langle\Delta r^2_{i, \rm CR}(t)\rangle_{\rm iso}$ at the relaxation time identify the locations more prone to structural rearrangements~\cite{Widmer-Cooper2004, Harrowell_NP}.

We investigate the correlations between $\langle\Delta r^2_{i, \rm CR}(t)\rangle_{\rm iso}$ and a local plastic length scale $\xi_i = \langle \Delta x_i(t_{\rm obs}) \rangle_\theta$, we define as the displacement induced by the application of a randomly oriented force of magnitude $f = 140$ on particle $i$, averaged over $200$ realizations.
To compare with previous works~\cite{Harrowell_NP,Patrick2014,Candelier2010}, we also investigate how $\langle\Delta r^2_{i, \rm CR}(t)\rangle_{\rm iso}$ correlates with the CR-DWF $\langle u_i^2\rangle = \langle\Delta r^2_{i, \rm CR}(\tau_\beta)\rangle_{\rm iso}$, the mode participation $p_i$~\cite{Harrowell_NP}, and the harmonic mean square displacement $\psi_i$~\cite{Patrick2014}.
Here $\tau_\beta$ is the time at which the logarithmic derivative of the mean square displacement acquires its minimum, implying that the system is maximally subdiffusive~\cite{Larini2008}.
To evaluate $p$ and $\psi$, we diagonalize the Hessian matrix of the disordered solid obtained by minimizing the energy of the $t = 0$ configuration via the Tinker package~\cite{Tinker}, obtaining $2N-2$ eigenmodes $\mathbf{e}(\omega_a)$ with non-zero eigenfrequencies $\omega_a$.
We define the participation fraction of particle $i$ as $p_i= \langle |\mathbf{e}_i(\omega_j)|^2\rangle_{N_m}$, where $j$ runs over the $N_m=50$ modes with lowest non-zero frequency.
The mean square displacement in the harmonic approximation is $\psi_i=\sum_{a}\omega_{a}^{-2}\left|\mathbf{e}_i(\omega_a)\right|^{2}$.
At each time $t$, we 
resort to Spearman's rank correlation coefficient $S(t)$~\cite{Sammut2011,Harrowell2006,tong_Tanaka,Reichman2014} to evaluate how $\langle\Delta r^2_{\rm CR}(t)\rangle_{\rm iso}$ correlates to each of the different considered static quantities.
$S=1$ ($-1$) for monotonically increasing (decreasing) dependence of two quantities, while $S=0$ in the absence of correlations.

\begin{figure}[t!]
 \centering
 \includegraphics[angle=0,width=0.48\textwidth]{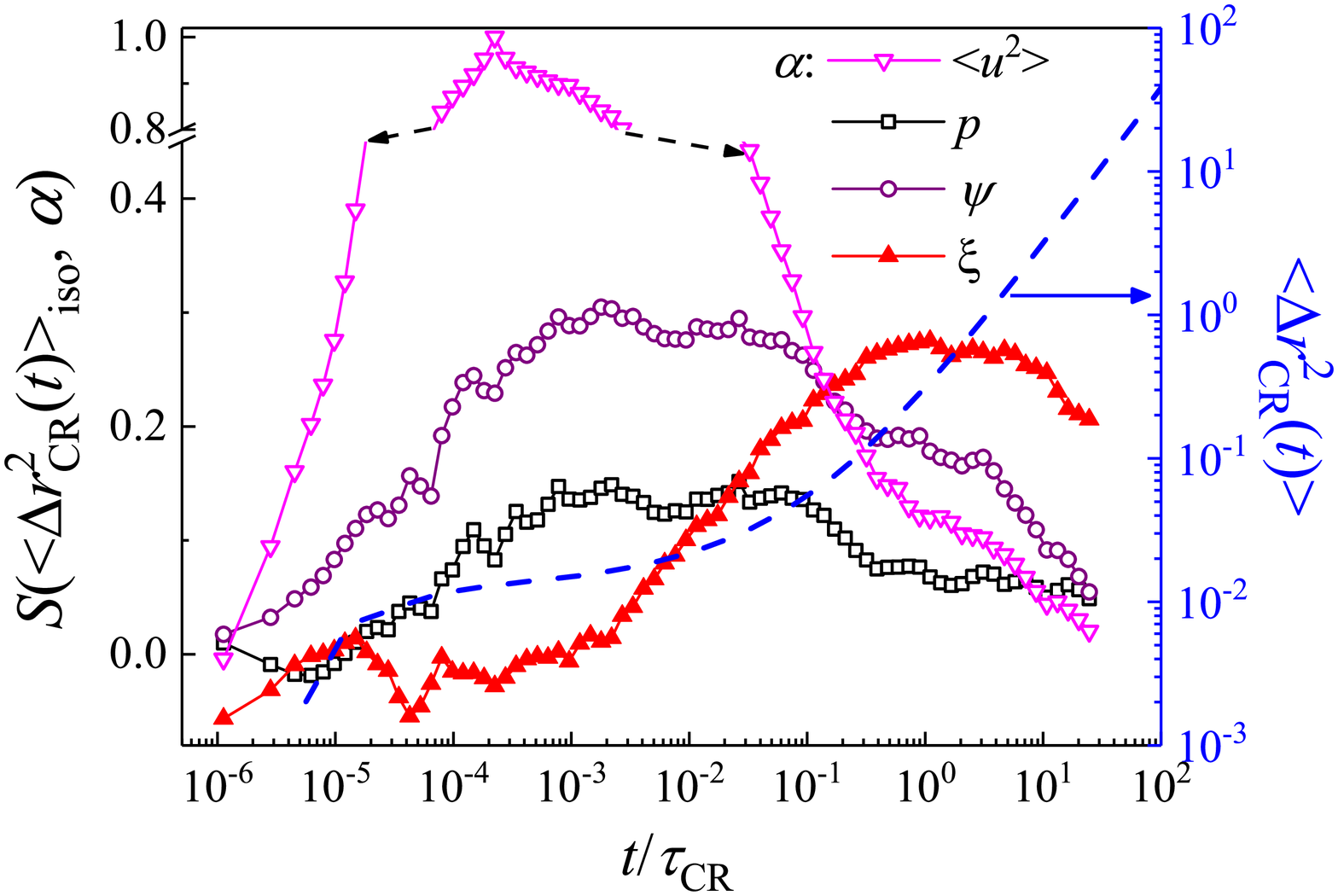}
 \caption{Time dependence of Spearman's rank correlation coefficient $S$ between the CR mean square displacement of the particles at time $t$ and
 1) their CR mean square displacement at $t = \tau_\beta$; 2) their participation fraction $p$; 3) their mean-square displacement evaluated in the harmonic approximation and 4) their plastic length scale $\xi$.
 The blue dashed line represents the average CR mean square displacement.
\label{fig:correlation}
}
\end{figure}

Figure~\ref{fig:correlation} illustrates (right axis) the time evolution of the average CR mean square displacement at $\rho = 1.2$ and $T = 0.4$, and (left axis) the time evolution of Spearman's rank correlation coefficients.
By definition, $\langle\Delta r^2_{i, \rm CR}(t)\rangle_{\rm iso}$ and $\langle u_i^2 \rangle$ are perfectly correlated at $t = \tau_\beta$ ($\tau_\beta \simeq 3\times10^{-4}\tau_{\rm CR}$ at $\rho = 1.2$ and $T = 0.4$).
Figure~\ref{fig:correlation} clarifies that the correlation sharply drops at longer times~\cite{Patrick2014}, equaling $\simeq 0.1$ at the relaxation time scale.
Similarly, the correlation between $\langle\Delta r^2_{i, \rm CR}(t)\rangle_{\rm iso}$ and both participation fraction and mean square displacement in the harmonic approximation reaches a maximum at short times and sensibly drops on the relaxation time scale.

These results agree with previous investigations~\cite{Harrowell_NP} on the relation between iso-configurational mean square displacement and participation fraction, that found correlations at $t \simeq 0.3\tau$, with $\tau$ the relaxation time evaluated without resorting to CR measures~\cite{Perera1999}.
Indeed, for the system considered in Fig.~\ref{fig:correlation}, $t = 0.3\tau$ corresponds to $t \simeq 0.1\tau_{\rm CR}$, a short time where the dynamics is still dominated by the vibrational motion.
The critical conclusion we reach from this investigation is that participation fraction and DWF correlate well with the vibrational dynamics but weakly with the dynamics on the relaxation time scale. 
Henceforth, these quantities are not a good proxy for the location of the relaxation events.

In Figure~\ref{fig:correlation}, we also illustrate Spearman's rank correlation coefficient between $\langle\Delta r^2_{i, \rm CR}(t)\rangle_{\rm iso}$ and the plastic length scale $\xi_i$.
This correlation coefficient is small at short times and reaches a maximum at the relaxation time. 
Hence, it critically differs from the other correlation coefficients we have investigated that peak on the vibrational time scale.
These findings demonstrate that the plastic length scale is a robust indicator of the relaxation dynamics at the particle level. 
This result is our second significant result.

To further test the validity of our results, we have replaced $\langle\Delta r^2_{i, \rm CR}(t)\rangle_{\rm iso}$ with alternative measures of structural relaxation, a particle-based self-scattering function and the fraction of neighbors the particle has lost at time $t$.
Fig~S11~\cite{SM} shows that our finding are robust with respect to the adopted definition of local structural relaxation.
In addition, in the SM we show that
the average plastic length scale correlates with the relaxation time, and the locally defined plastic length scale identifies the particles more prone to rearrange,
also when the dynamical slowdown is induced by an increase of the density at a constant temperature.
These investigations support the robustness of our findings and clarify that our predictions could be experimentally tested in suspensions of colloidal particles~\cite{GranickNature2020, Weeks_linearElastic, Lucio_PNAS, Jasper_SA}, where density is the main control parameter.

Our results show that the response of individual particles to transient, almost impulsive perturbations defines a plastic correlation length that regulates supercooled liquids' relaxation dynamics.
The average plastic correlation length $\xi$ is inversely proportional to the free energy barrier $\Delta F(T)$ governing the dynamic slowdown.
The plastic response of individual particles informs on their participation in localized relaxation events, and hence on the spatial heterogeneity of the relaxation process at the relaxation time scale.
Our findings provide novel insights into the origin of slow and heterogeneous dynamics, showing that the local elastic response mostly correlates with the vibrational dynamics rather than with the dynamics at the relaxation time. 
Our results may inspire novel glass transition theories and stimulate experimental research on the plastic response to external forces induced, e.g., by laser pulses.

We thank M. Wyart for fruitful discussions.
Y.-W.L. acknowledges support of the start-up funding of Beijing Institute of Technology and support
from the National Natural Science Foundation (NSF) of China (Grants No. 12105012). Y.Y. is supported by the NSF of China (Grants Nos. 11734003, 12061131002).
MPC acknowledges support from the Singapore Ministry of Education through the Singapore Academic Research Fund (MOE2019-T1-001-03 \& MOE2019-T1-001-22), and the National Supercomputing Centre Singapore (NSCC) for the computational resources.

\bibliographystyle{apsrev4-1}

%
\end{document}


\title{Supplemental Material \\ for \\
  Local Plastic Response and Slow Heterogeneous Dynamics of Supercooled Liquids
}

\author{Yan-Wei Li}
\affiliation{Key Laboratory of Advanced Optoelectronic Quantum Architecture and Measurement (MOE), School of Physics, Beijing Institute of Technology, Beijing, 100081, China}
\author{Yugui Yao}
\affiliation{Key Laboratory of Advanced Optoelectronic Quantum Architecture and Measurement (MOE), School of Physics, Beijing Institute of Technology, Beijing, 100081, China}
\author{Massimo Pica Ciamarra}
\affiliation{Division of Physics and Applied Physics, School of Physical and
Mathematical Sciences, Nanyang Technological University, Singapore 637371, Singapore}
\affiliation{
CNR--SPIN, Dipartimento di Scienze Fisiche,
Universit\`a di Napoli Federico II, I-80126, Napoli, Italy
}
\affiliation{
CNRS@CREATE LTD, 1 Create Way, \#08-01 CREATE Tower, Singapore 138602
}
\maketitle

\onecolumngrid
\tableofcontents
~\newpage
\section{Slow dynamics and onset temperature\label{sec:dyn}}

\begin{figure}[h!]
 \centering
 \includegraphics[angle=0,width=0.45\textwidth]{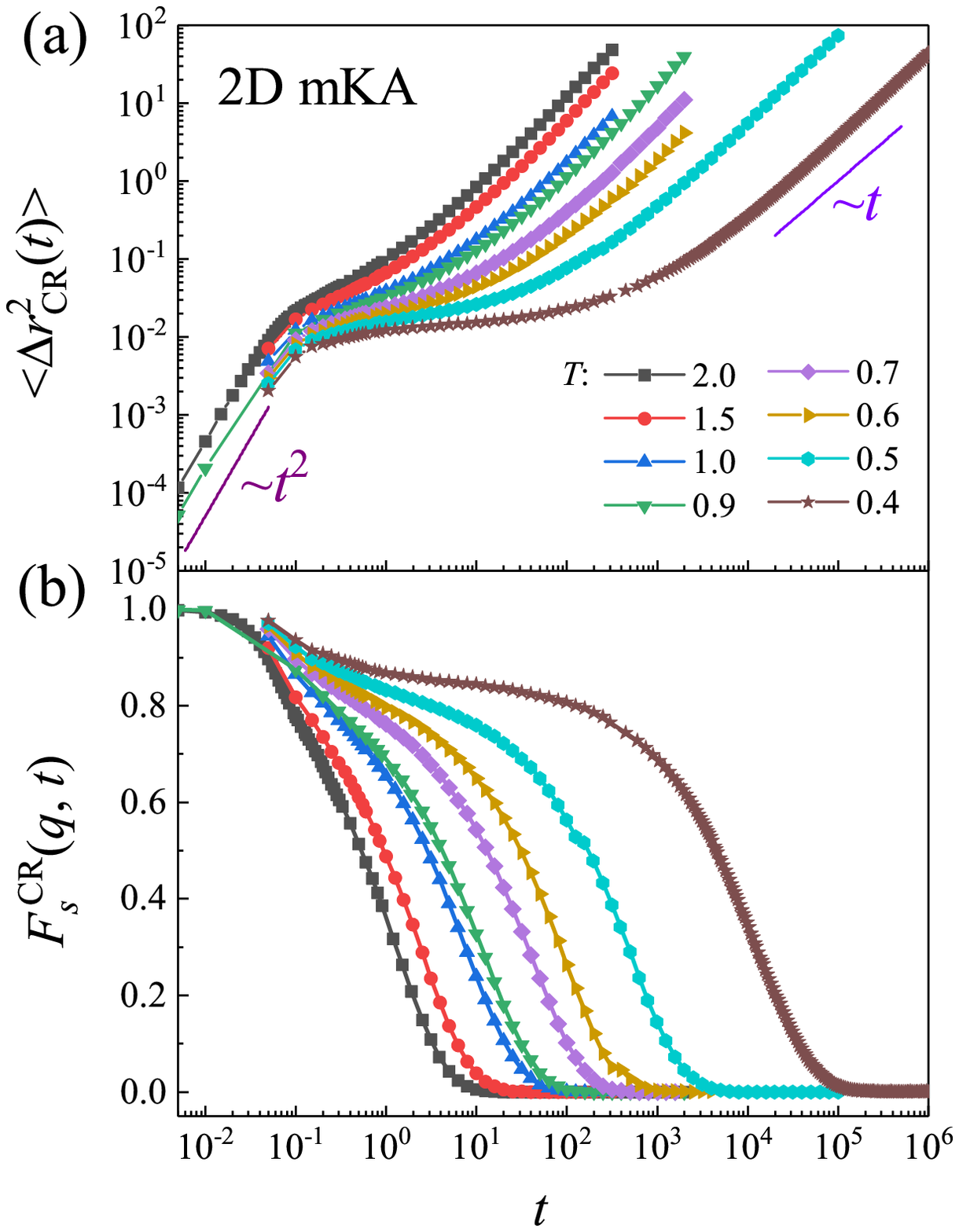}
 \caption{Time dependence of the cage-relative mean square displacement (a), and of the cage-relative intermediate scattering function (b) for different temperatures as shown in the legend of (a).
\label{fig:msdisf}
}
\end{figure}
The slow-dynamics of the 2D Kob-Andersen mixture we have considered has been previously studied~\cite{mKA,Yanwei_MassimoPNAS,Shiba}.
Here we recap its main features for sake of completeness, illustrating the temperature dependence of the cage-relative mean square displacement (CR-MSD) and the cage-relative self-intermediate scattering function (CR-ISF). 
The CR-MSD is defined as $\langle \Delta r_{\rm CR}^{2}(t)\rangle=\langle \frac{1}{N}
\sum_{i=1}^{N}\Delta \mathbf{r}_{i}^{\rm CR}(t)^{2}\rangle$, where $\Delta \mathbf{r}_{i}^{\rm CR}(t)$ is the cage-relative displacement of particle $i$, $\Delta \mathbf{r}_{i}^{\rm CR}(t) =\mathbf{r}_{i}(t)-\mathbf{r}_{i}(0) - 1/N_{i}\sum_{j=1}^{N_{i}}[\mathbf{r}_{j}(t)-\mathbf{r}_{j}(0)]$ with $N_{i}$ the number of neighbors of particle $i$ evaluated at time $0$. 
The CR-ISF is $F_{s}^{\rm CR}(q,t)=\langle \frac{1}{N} \sum_{j=1}^{N}e^{i\mathbf{q}\cdot\Delta \mathbf{r}_{i}^{\rm CR}(t)}\rangle$ with $q=|\mathbf{q}|$ the wavenumber of the first peak of the static structure factor. 
Figure~\ref{fig:msdisf} illustrates the appearance of a transient solid-like response as the temperature decrease.

\begin{figure}[h!]
 \centering
 \includegraphics[angle=0,width=0.45\textwidth]{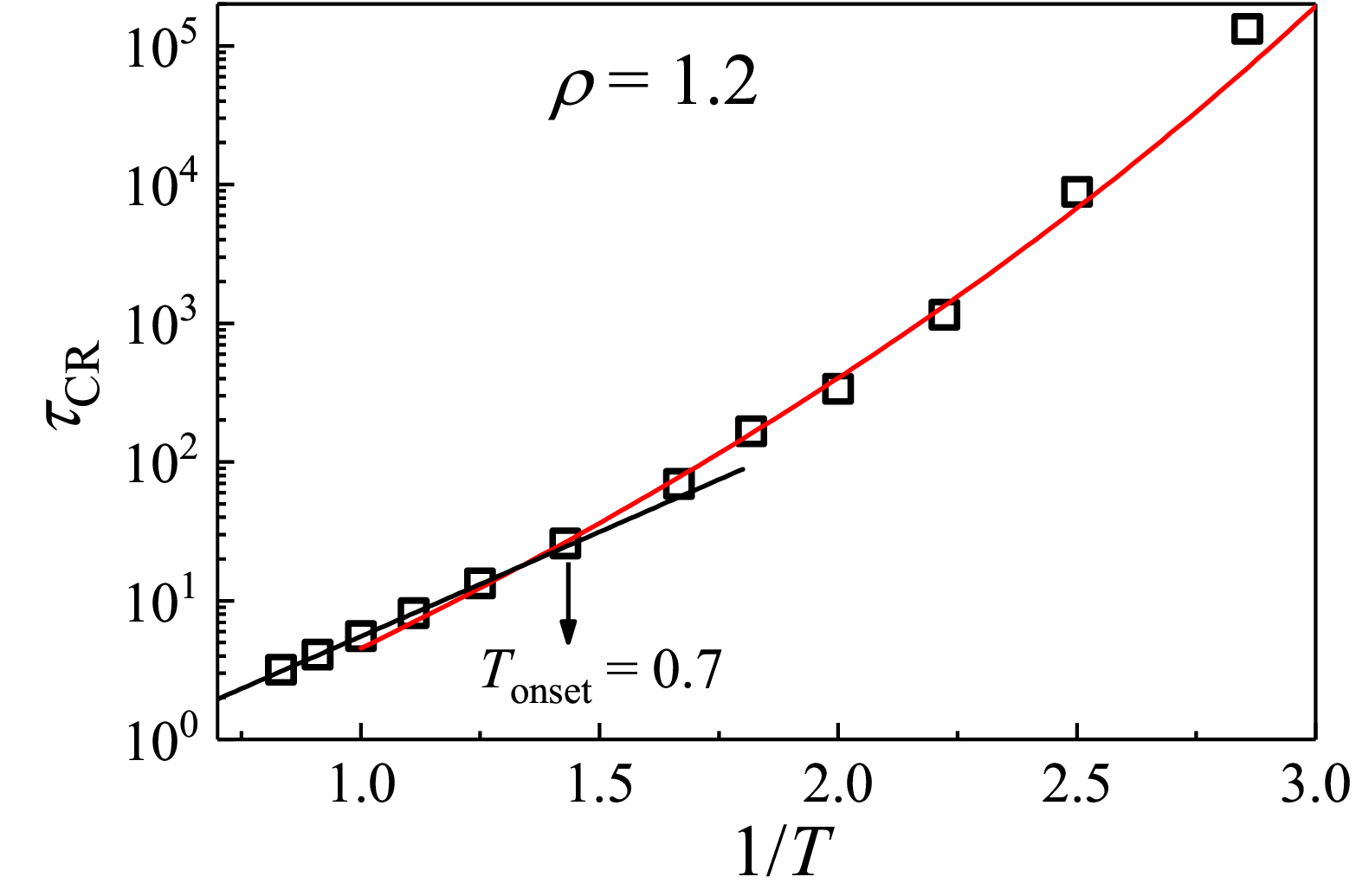}
 \caption{$T$ dependence of the CR relaxation time. $\tau_{\rm CR}$. The black and red lines are Arrhenius and VFT fittings, respectively.
\label{fig:ArrVFT}
}
\end{figure}

We extract the cage-relative relaxation time from the decay of the self-scattering function, $F_{s}(q,\tau_{\rm CR}) = 1/e$, and illustrate its temperature dependence in Fig.~\ref{fig:ArrVFT}. 
We observe a crossover from a high temperature Arrhenius behavior (black line) to a low-temperature super-Arrhenius regime well described by the Vogel-Fulcher-Tammann (VFT) equation (red line)
$\tau_{\rm CR}=\tau_0 \exp\left(B/(T-T_0)\right)$ with $T_0=0.12$ and $B/T_0=25$. 
These two regimes meet at the onset temperature $T_{\rm onset}=0.7$. 
The value of the characteristic VFT temperature $T_0$ is consistent with that where the plastic length scales extrapolates to zero in Fig. 1(b) of the main text. Data refer to $\rho  = 1.2$.

\begin{figure}[!!h]
\centering
\includegraphics[width=0.55\textwidth]{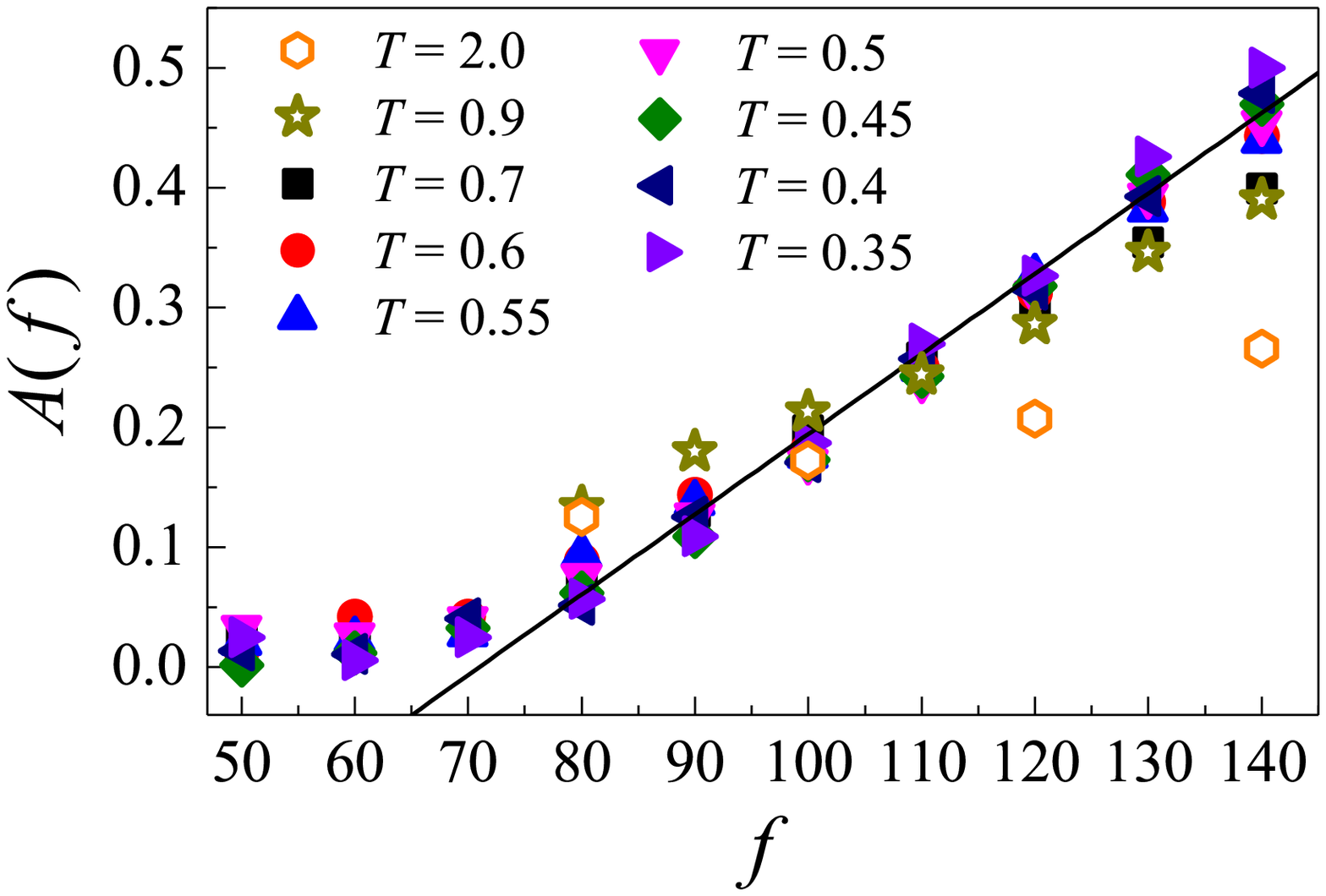}
 \caption{
 $A(f) = L(T,f)/(T-T_0)$ as a function of $f$, both above and below the onset temperature, $T_{onset}=0.7$.
\label{fig:Afall}
}
\end{figure}

The above onset crossover at $T_{\rm onset}=0.7$ indicates that the observed scaling at $T<T_{\rm onset}$ in Figs. 1(b) and 1(c) of the main text may break at higher temperatures. We illustrate in Fig.~\ref{fig:Afall} the same plot as Fig. 1(c) of the main text in addition to the high $T$ data, and find deviations of high $T$ data from the low $T$ ones. 

~\newpage
\section{Plastic length distribution\label{sec:distribution}}
\begin{figure}[h!]
 \centering
 \includegraphics[angle=0,width=0.45\textwidth]{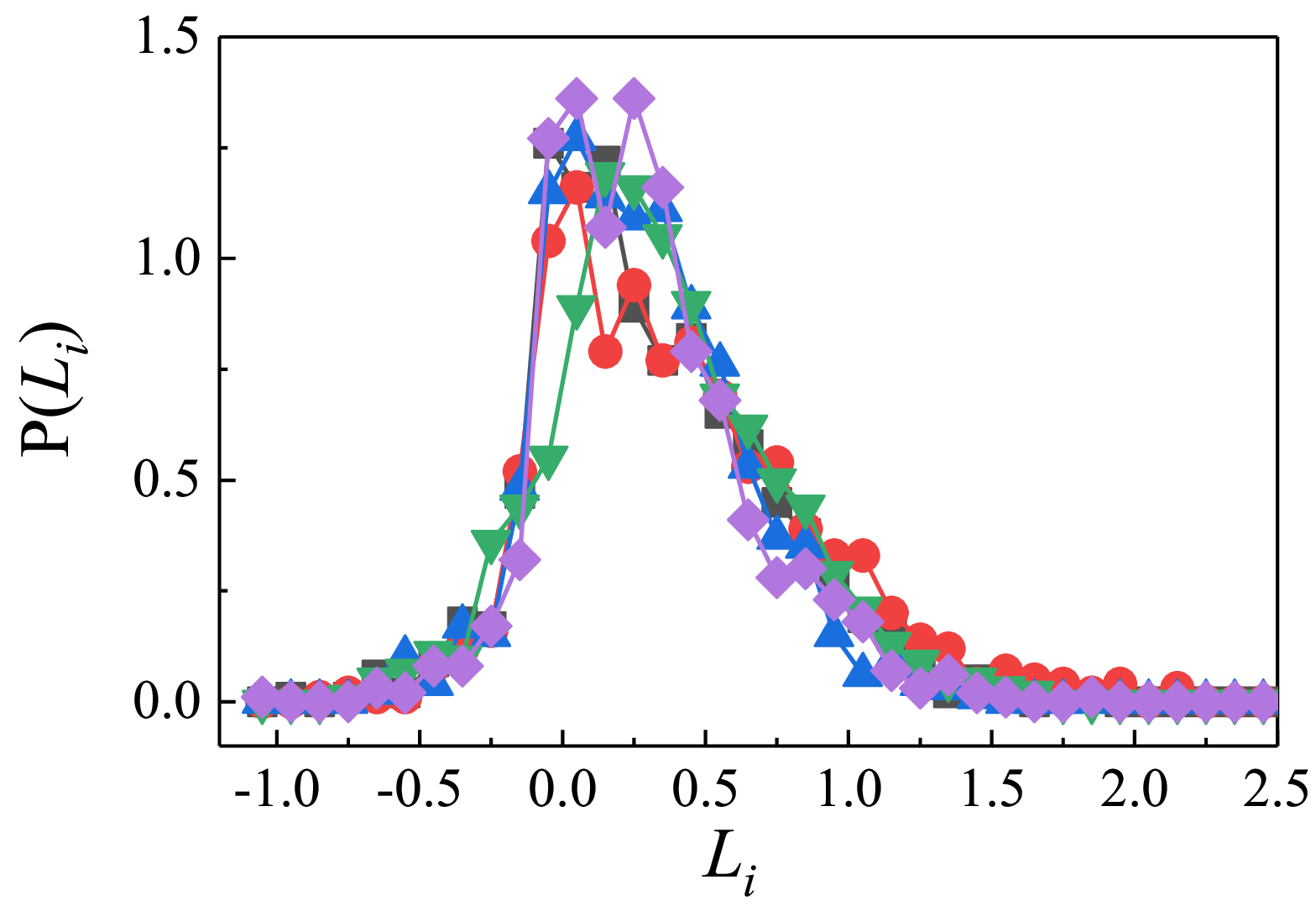}
 \caption{Probability distribution of the displacement $L_i$ of particle $i$ induced by a transient force.
Different colours identify different particles.
We evaluate each distribution by applying 1000 randomly oriented forces of magnitude $f=140$ on each particle.  Data refer to $\rho=1.2$ and $T=0.4$.
\label{fig:pdf}
}
\end{figure}

A particle undergoes a plastic displacement in response to a transient force of magnitude $f$. 
We associate a set of displacements to each particle by randomly changing the direction of the applied force. Fig.~\ref{fig:pdf} illustrates the probability distribution of these displacements for five randomly selected particles. 
The distributions are asymmetric as particles preferentially move in the direction of the applied force. The skewness is $\simeq 0.15$.
~\newpage
\section{Role of damping parameter, force duration and force magnitude\label{sec:parameters}}
\begin{figure}[h!!]
\centering
\includegraphics[width=0.45\textwidth]{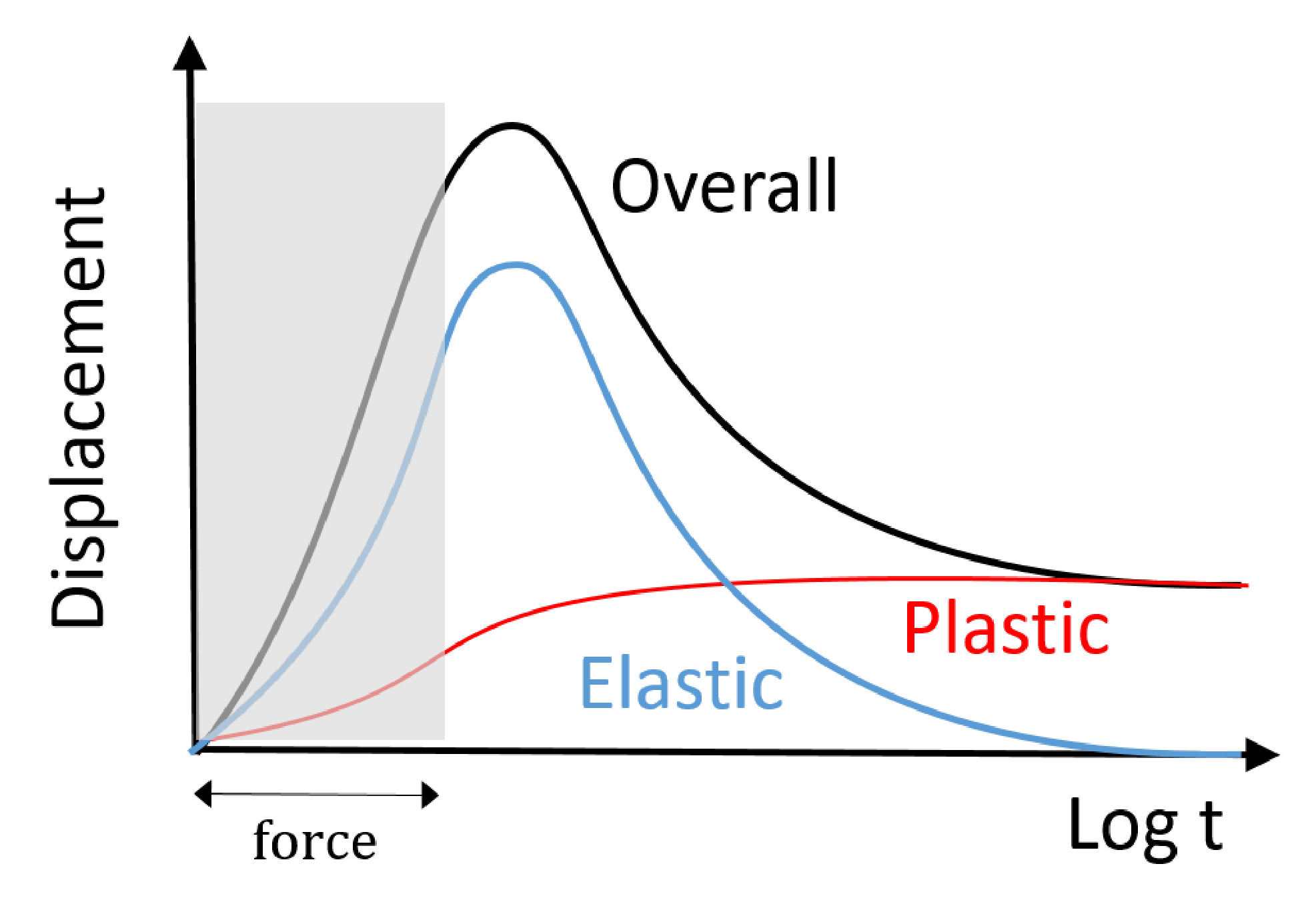}
\caption{
Schematic decomposition of the overall (black line) induced displacement in its plastic (red line) and elastic (light blue line) components.
\label{fig:schematic}
}
\end{figure}

We use a transient perturbation to displace a particle.
At short time, the induced displacement comprises an elastic and an irreversible component.
At longer times, the elastic displacement retracts to zero, so that the overall induced displacement reduces to the irreversible part, as schematically illustrated in Fig.~\ref{fig:schematic}.
With this approach, we induce and characterize transitions between adjacent minima in the system's free energy landscape.

Three parameters influence the response of a particle to the transient force: the friction coefficient $\gamma$ in the Langevin equation, the magnitude of the force $f$ and the force duration $t_{\rm per}$. 
In the following we study the role of these three parameters on the induced displacement.
We show that our results are robust if these parameters are chosen so that the maximum value of the induced displacement is a fraction of the typical interparticle distance, to ensure the occurrence of a plastic response.
\begin{figure*}[h!!]
\centering
\includegraphics[width=0.85\textwidth]{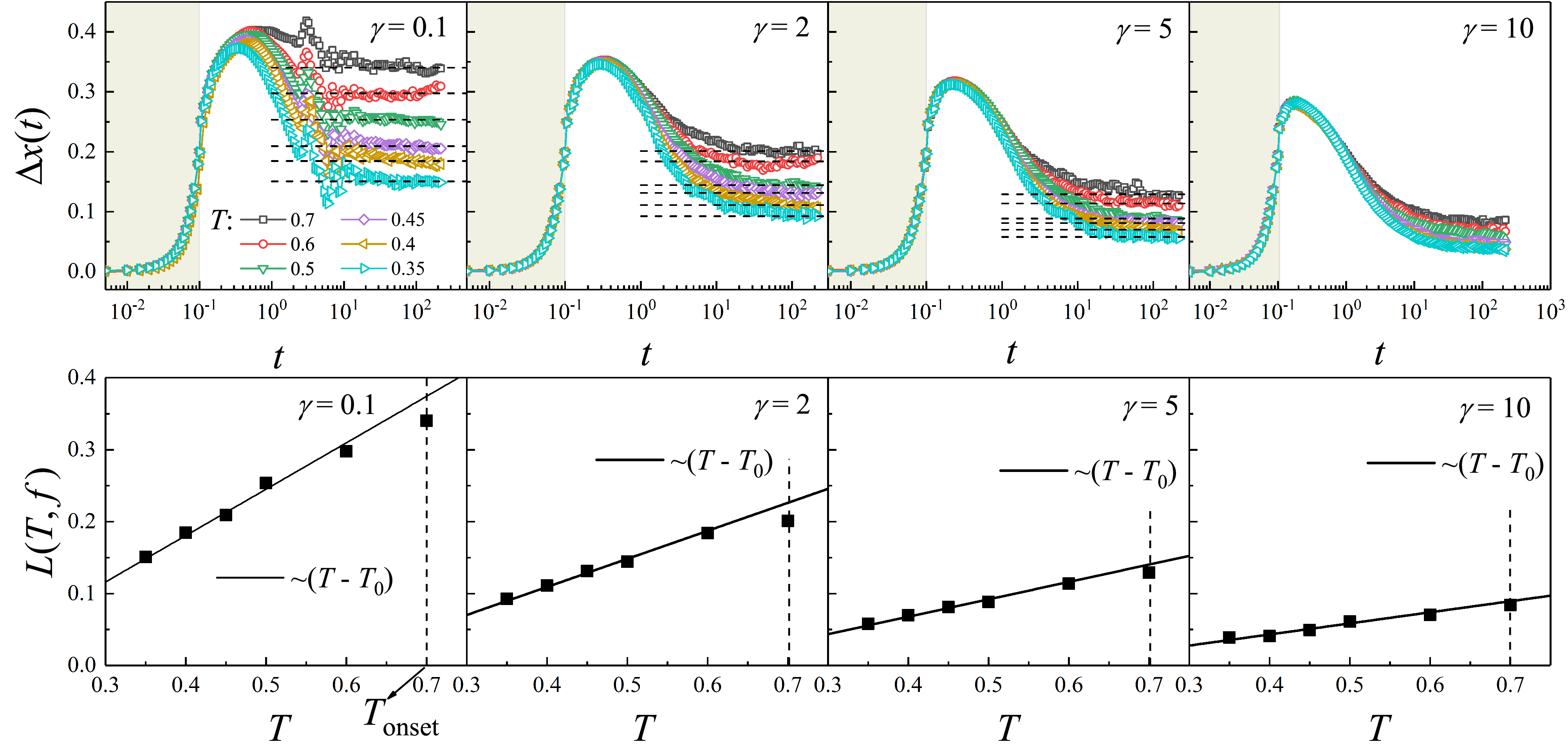}
\caption{
The top row illustrates the average displacement induced by a transient perturbation with $f=140$ as the damping parameter increases (from left to right).
The bottom row illustrate the dependence of the force induced displacement on the temperature.
We find $L(T, f=140)\propto (T-T_0)$ for different $\gamma$, as in the main text.
\label{Fig:gamma}
}
\end{figure*}

Fig.~\ref{Fig:gamma} illustrates the time dependence of the displacement $\Delta x(t)$ for different values of $\gamma$ (top row), and the corresponding temperature dependence of the induced irreversible displacement $L$ (bottom row).
In these figures, the applied force has a constant magnitude $f=140$ and duration $t_{\rm per}=0.1$, while the damping parameter increases from left to right.
The maximum value of $\Delta x(t)$ and the induced irreversible displacement $L$ reduce as $\gamma$ increases. The scaling $L\propto(T-T_0)$ clearly holds for $\gamma\leq10$.
At larger $\gamma$ values the scaling breaks as damping prevents the occurrence of relevant irreversible events.

\begin{figure*}[h!!]
\centering
\includegraphics[width=0.5\textwidth]{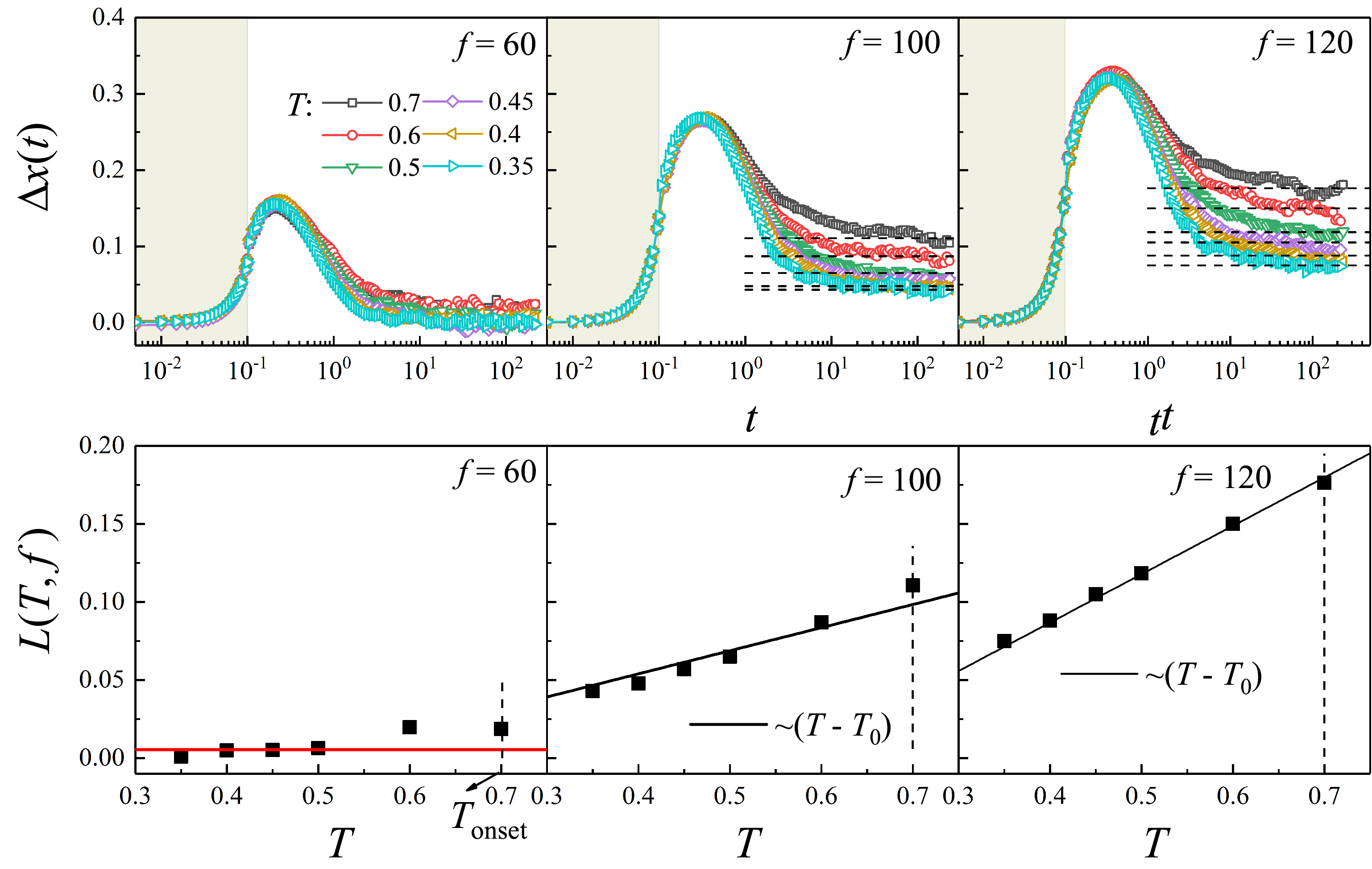}
\caption{
The top row illustrates the average displacement induced by the transient perturbation with damping parameter $\gamma=1$, as the magnitude of force increases (from left to right).
The bottom row illustrates the dependence of the force induced displacement on the temperature.
\label{Fig:force}
}
\end{figure*}

To recover the correct scaling at large $\gamma$ one needs to increase the force magnitude (or duration).
We show that this is indeed the case by investigating the induced displacement's dependence on the magnitude of the applied force.
Fig.~\ref{Fig:force} shows that if at a given force magnitude $f$ the scaling $L\propto(T-T_0)$ does not hold, this is recovered on increasing $f$.

\begin{figure}[h!!]
\centering
\includegraphics[width=0.48\textwidth]{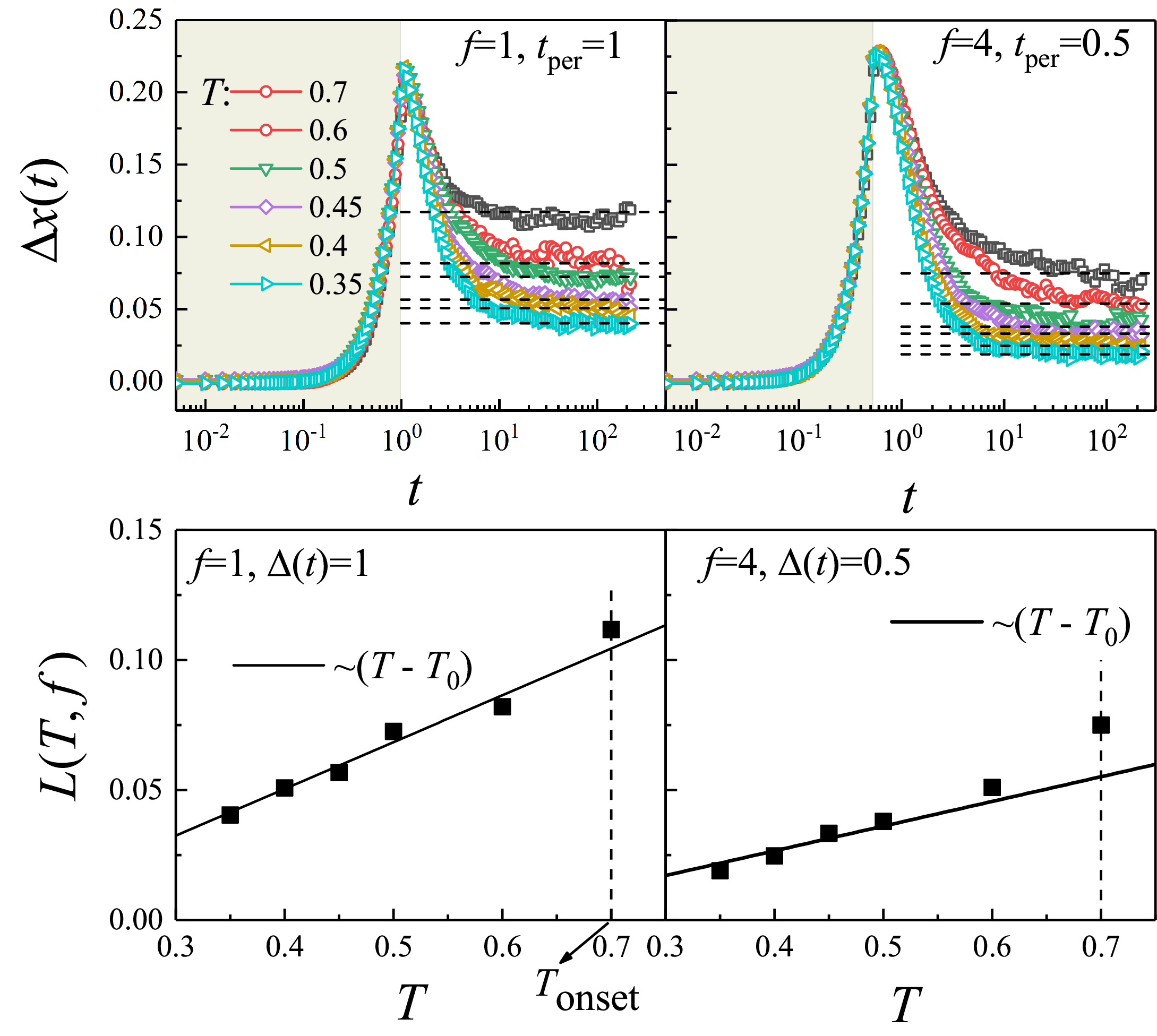}
 \caption{
 The scaling $L\propto(T-T_0)$ is robust to changes in the force duration, we generally fix to $t_{\rm per} = 0.1$.
 If the duration of the perturbation increases, the scaling is recovered provided the force magnitude is adjusted so that the maximum induced displacement is a fraction of the typical interparticle distance.
\label{Fig:Fduration}
}
\end{figure}

In Fig.~\ref{Fig:Fduration} we further show that the scaling is robust with respect to variation in the force duration $t_{\rm per}$ if the force magnitude is adjusted to ensure the occurrence of irreversible events.

To summarise, we find that the three parameters, $\gamma$, $f$, and $t_{\rm per}$ influence the induced displacement. 
As a rule of thumb, we find that our scaling works if the maximum induced displacement is at least 0.2 interparticle spacing. This is of the order of the typical displacement threshold used in the literature to identify irreversible events.

~\newpage
\section{Plastic response at $T = 0$\label{sec:T0}}
At finite temperature, the plastic displacement induced by the transient application of a force of magnitude $f$ on a particle vanishes if the force is smaller than a temperature-independent threshold $f_0$, as illustrated in Fig.~1(c) of the main text.
For $f > f_0$, the induced displacement is proportional to $f-f_0$.
We rationalize the temperature independence of $f_0$ and the linear $f$ dependence of the induced displacement by investigating the response of individual particles to localized forces in the zero-temperature limit.

We use the following protocol. 
First, we minimize the energy of a system with an initial (parent) temperature $T_p$, bringing it into its closest inherent structure (IS).
Then, we apply a force of magnitude $f$ to a randomly selected particle in a randomly chosen direction $\hat f$. We ensure the net force applied to the system vanishes to prevent the motion of the centre of mass by also exerting forces of magnitude $f/(N-1)$ and direction $-\hat f$ to all other particles.
We quasistatically increase the magnitude of $f$ up to a value $f_{\rm max}$ and then decreases it back to zero while keeping track of the displacement of the perturbed particle, $l_{\rm IS}(f)$.
\begin{figure}[h!]
 \centering
 \includegraphics[angle=0,width=0.45\textwidth]{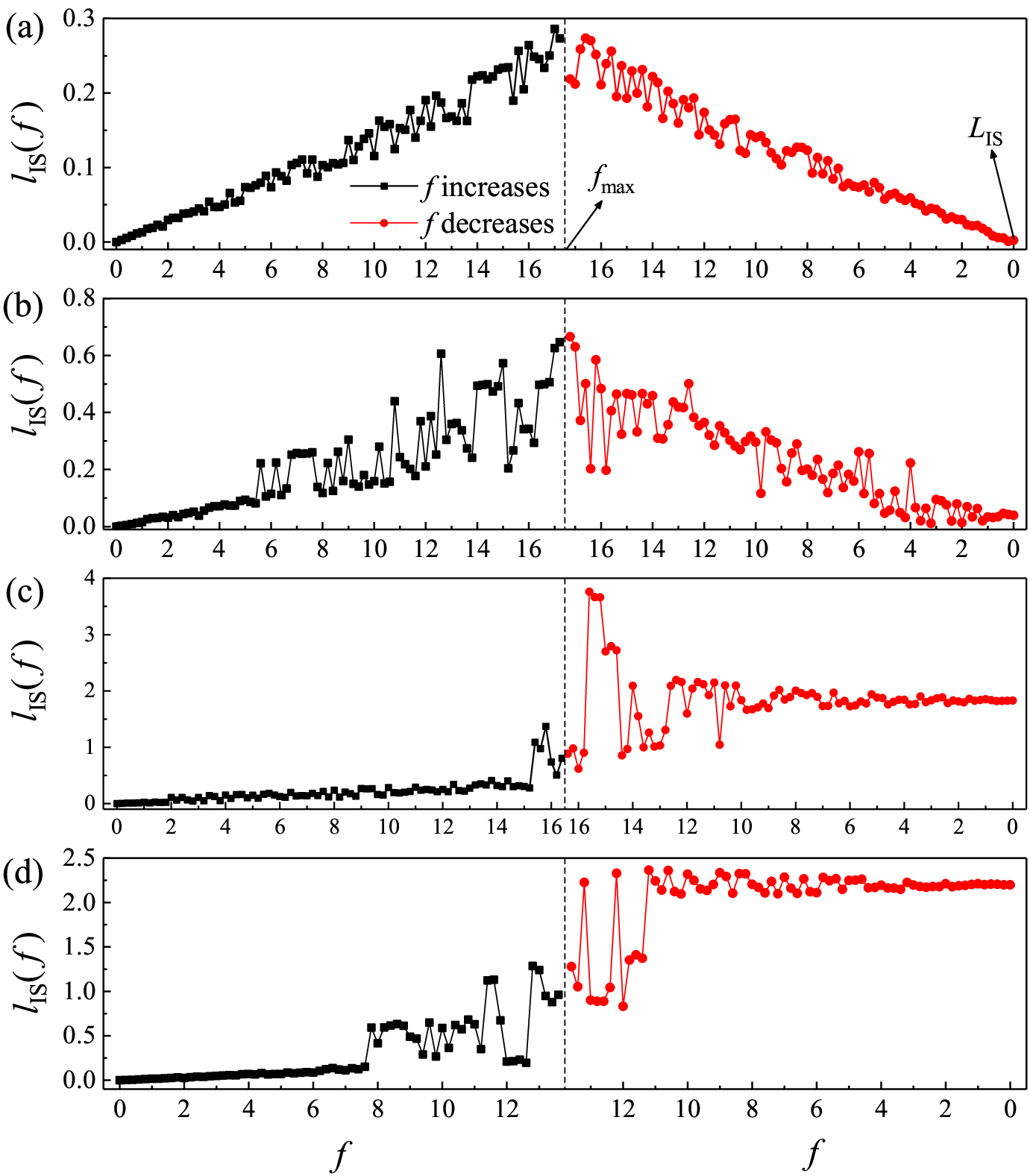}
 \caption{Evolution of the displacement of a particle upon the application of a force $f$ that first grows (black squares) up to $f_{\rm max}$ and then decrease (red circles) to zero.
 The final displacement at the end of the force cycle defines the inherent structure length, $L_{\rm IS}$. 
 The panels illustrate results for different particles. 
 The parent temperature is $T_{\rm p}=0.35$.
\label{fig:lf}
}
\end{figure}

Fig.~\ref{fig:lf} illustrates how the induced displacement varies as the magnitude of applied force increases up to $f_{\rm max}$ and then decreases back to zero, for four particles.
The granular increase of the displacement is interrupted by sudden jumps.
In some cases, e.g., panel (a), these jumps are reversible, as the displacement decreases and essentially vanishes at the end of the force cycle. 
In other cases, jumps correspond to plastic rearrangements which are not reversed as the force decreases, as in panels (b)-(d).
As such, the displacement of the particle at the end of the force cycle defines an IS-length $L_{\rm IS}(T_{\rm p},f_{\rm max})$ that relates to the induced plastic response.
\begin{figure}[!!t]
 \centering
 \includegraphics[angle=0,width=0.45\textwidth]{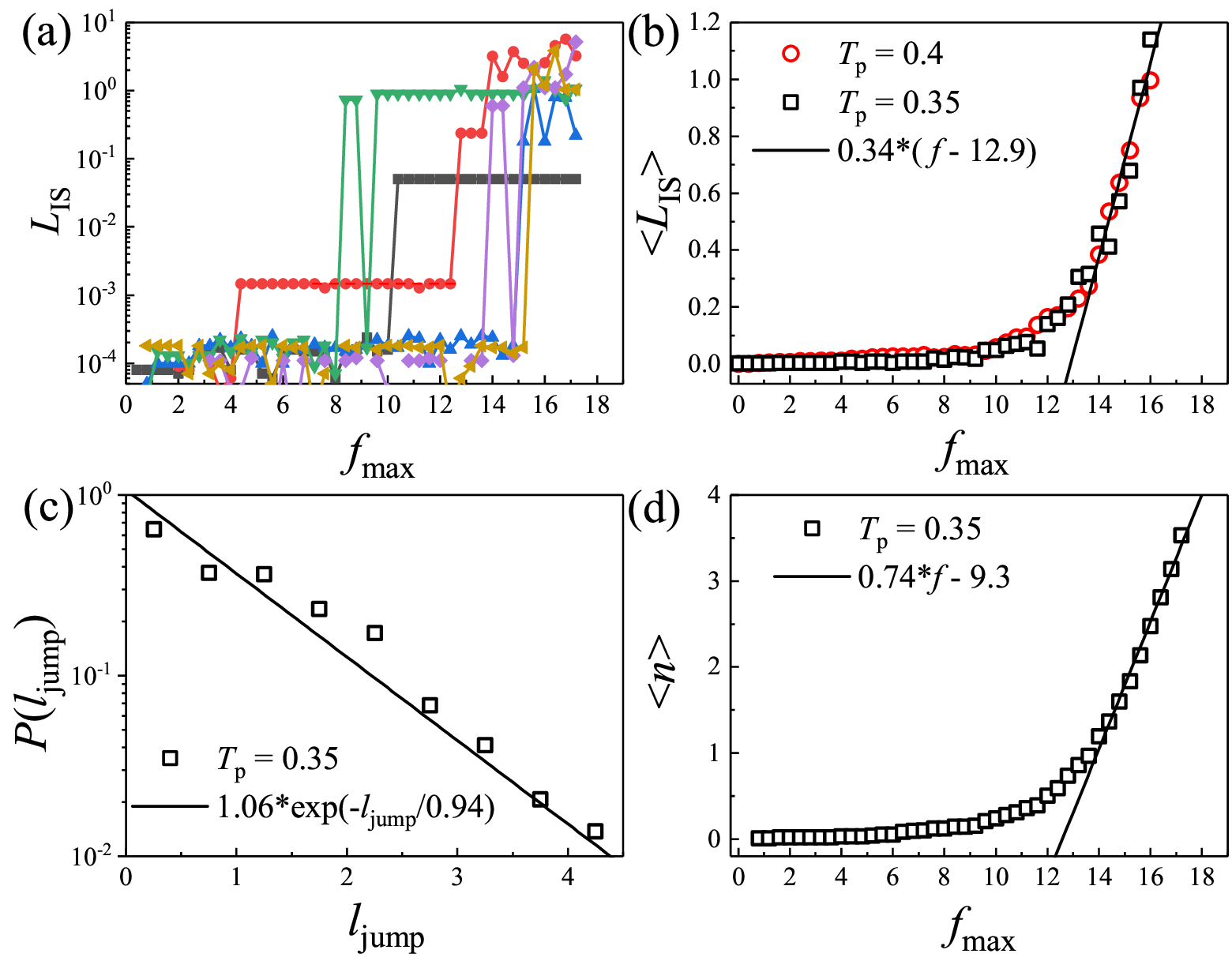}
 \caption{(a) Dependence of the IS-length $L_{\rm IS}$ of a few selected-particles on the maximum force reached during the force cycle, at parent temperature $T_p = 0.35$.
 The IS-length grows via a sequence of jumps.
 (b) Dependence of the average IS-length $\langle L_{\rm IS}\rangle$ on $f_{\rm max}$. 
 (c) Probability distribution of the magnitude of the displacements jumps $l_{\rm jump}$ seen in (a). 
 (d) Average number of jumps per particle the plastic jump event as a function of $f_{\rm max}$.
\label{fig:Lis}
}
\end{figure}

Fig.~\ref{fig:Lis}(a) illustrates the dependence of the IS-length on the maximum applied force, for a few particles.
The IS-length grows with $f_{\rm max}$ through a succession of jumps, each relating to the occurrence of a plastic event.
By averaging these data over different particles, we determine the dependence of the average IS-length $\langle L_{\rm IS} \rangle$ on $f_{\rm max}$.
$\langle L_{\rm IS} \rangle$ vanishes at small $f_{\rm max}$ values not able to trigger plastic rearrangements and grows linearly with $f_{\rm max}$ at large ones.
Results are largely temperature independent.

The force dependence of the IS-length $\langle L_{\rm IS} \rangle$ parallels that of the finite-temperature plastic length illustrated in Fig.~1(c) of the main text.
This result suggests that the minimum force $f_0$ above which finite-temperature systems respond plastically, and its temperature independence, reflect features of the underlying energy landscape. 
The actual values of the minimum force able to trigger plastic events for the IS and $T>0$ dynamics differ, as in the latter case, the force is only transiently applied and viscous forces influence the dynamics.

We further rationalize the linear dependence of the plastic length on the force by investigating the dependence of the IS-length $\langle L_{\rm IS} \rangle$ on $f_{\rm max}$.
Fig.~\ref{fig:Lis}(a) illustrates the plastic length scale evolves via a sequence of jumps. 
Using a standard thresholding procedure, we have identified these jumps and their length, $l_{\rm jump}$.
Fig.~\ref{fig:Lis}(c) shows that the jump length magnitude has an exponential distribution with average value $\langle l_{\rm jump} \rangle \simeq 0.94$.
In Fig.~\ref{fig:Lis}(d), we demonstrate that the average number of jumps varies with $f_{\rm max}$ as the IS-length.
These results suggest that the IS-length is $L(f_{\rm max})= \langle l_{\rm jump}\rangle n(f_{\rm max}) $, and that its force dependence reflects the probability that a perturbation induces an irreversible rearrangement.

Similarly, we expect the number of plastic events to control the linear dependence on $f$ of the $T > 0$ plastic length scale.

~\newpage
\section{Alternative measure of structural relaxation\label{sec:correlation}}
In the main text, we have evaluated the correlation between particle-based quantities defined on a short time scale, $\langle u^2\rangle$, $p$, $\Psi$ and $\xi$, and a particle's propensity as evaluated by the isoconfigurational MSD, $\langle\Delta r^2_{j, \rm CR}(t)\rangle_{\rm iso}$.

\begin{figure}[!!h]
\centering
\includegraphics[width=0.5\textwidth]{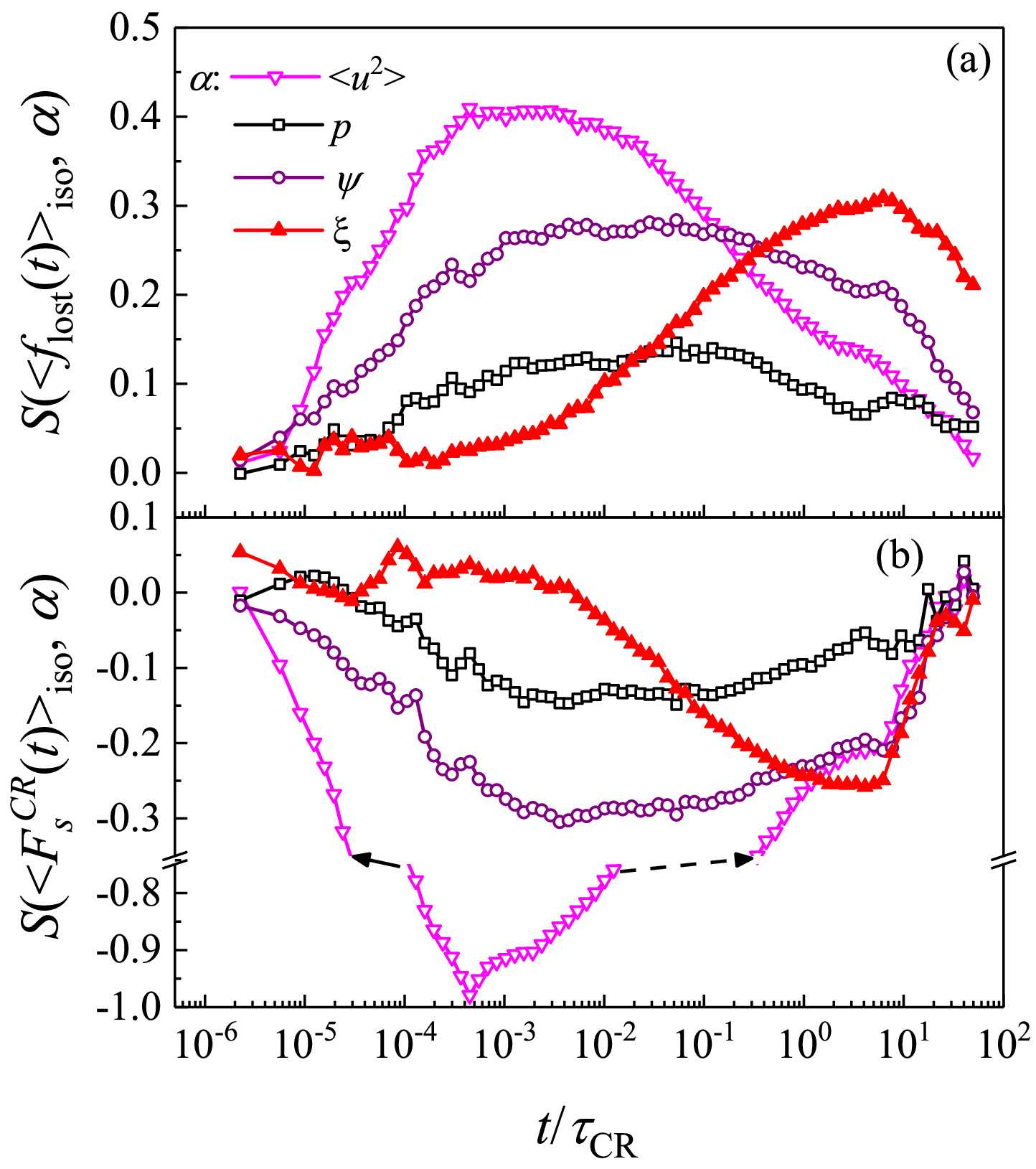}
\caption{
We find similar results as in Fig. 3 studying correlations with (a) the fraction of lost neighbors, $f_{\rm lost}$ and (b) the cage-relative intermediate scattering functions, $F_s^{CR}$.
\label{Fig:fraction}
}
\end{figure}

Here, we use two alternative measures to evaluate how likely a particle is to relax:
\begin{itemize}
\item[(i)] we associate to each particle $j$ the fraction of its $t=0$ Voronoi neighbors lost at time $t$. 
Specifically, we consider the quantity
$$f_{\rm lost,j}(t) = \langle 1 - \frac{\sum_i {\rm vn}_{(j,i,t)}{\rm vn}_{(j,i,0)}}{\sum_i {\rm vn}_{(j,i,0)}}\rangle_{\rm iso},$$
where ${\rm vn}_{(i,j,t)} = 1 (0)$ if particles $i$ and $j$ are (are not) Voronoi neighbors at time $t$. 
\item[(ii)] we associate to each particle $j$ its CR-ISF $$F_{s, j}^{\rm CR}(t)=\langle e^{i\mathbf{k}\cdot \Delta \mathbf{r}_{j}^{\rm CR}(t)} \rangle_{\rm iso},$$ with $\mathbf{k}$ the first peak of the static structure factor. 
\end{itemize}
The use of these alternative variables leads to Fig.~\ref{Fig:fraction}, which qualitatively reproduces Fig. 3 in the main text (in Fig.~\ref{Fig:fraction}(b) the correlation coefficient is negative as particles with a smaller CR-ISF are those that rearrange). 
This finding proves that our results are robust with respect to the adopted measure of structural relaxation.

~\newpage
\section{Density dependence\label{sec:general}}
\begin{figure}[h!]
 \centering
 \includegraphics[angle=0,width=0.45\textwidth]{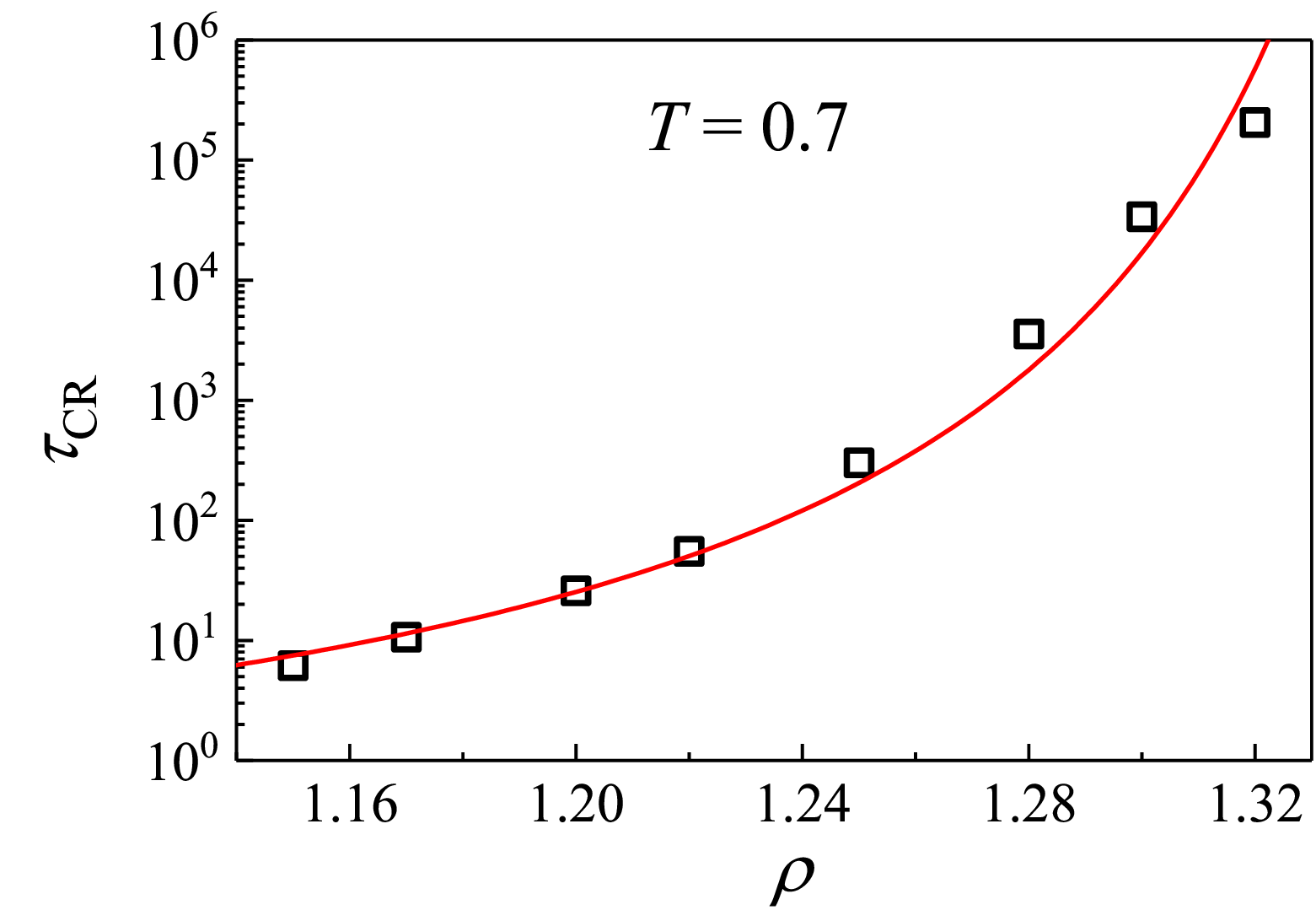}
 \caption{$\rho$ dependence of $\tau_{\rm CR}$ at $T=0.7$. The red line is a VFT fitting.
\label{fig:vft_rho}
}
\end{figure}

In the main text, we have demonstrated that the plastic length scale $\xi$ describes the slowdown of the dynamics that occurs as the temperature decreases at constant density.
To test the generality of our finding, here we investigate the slowing down of the dynamics induced by an increase in the density $\rho$ at constant temperature $T = 0.7$.
Fig.~\ref{fig:vft_rho} shows that the cage-relative relaxation time $\tau_{\rm CR}$ increases with the density.
The observed density dependence is reasonable described by VFT functional form $\tau_{\rm CR}=\tau_0{\rm exp}(B/(\rho_0-\rho))$, where $\rho_0 = 1.39$, $B/T_0=0.8$, and $\tau_0=0.073$, in the considered density range.

\begin{figure}[h!]
 \centering
 \includegraphics[angle=0,width=0.45\textwidth]{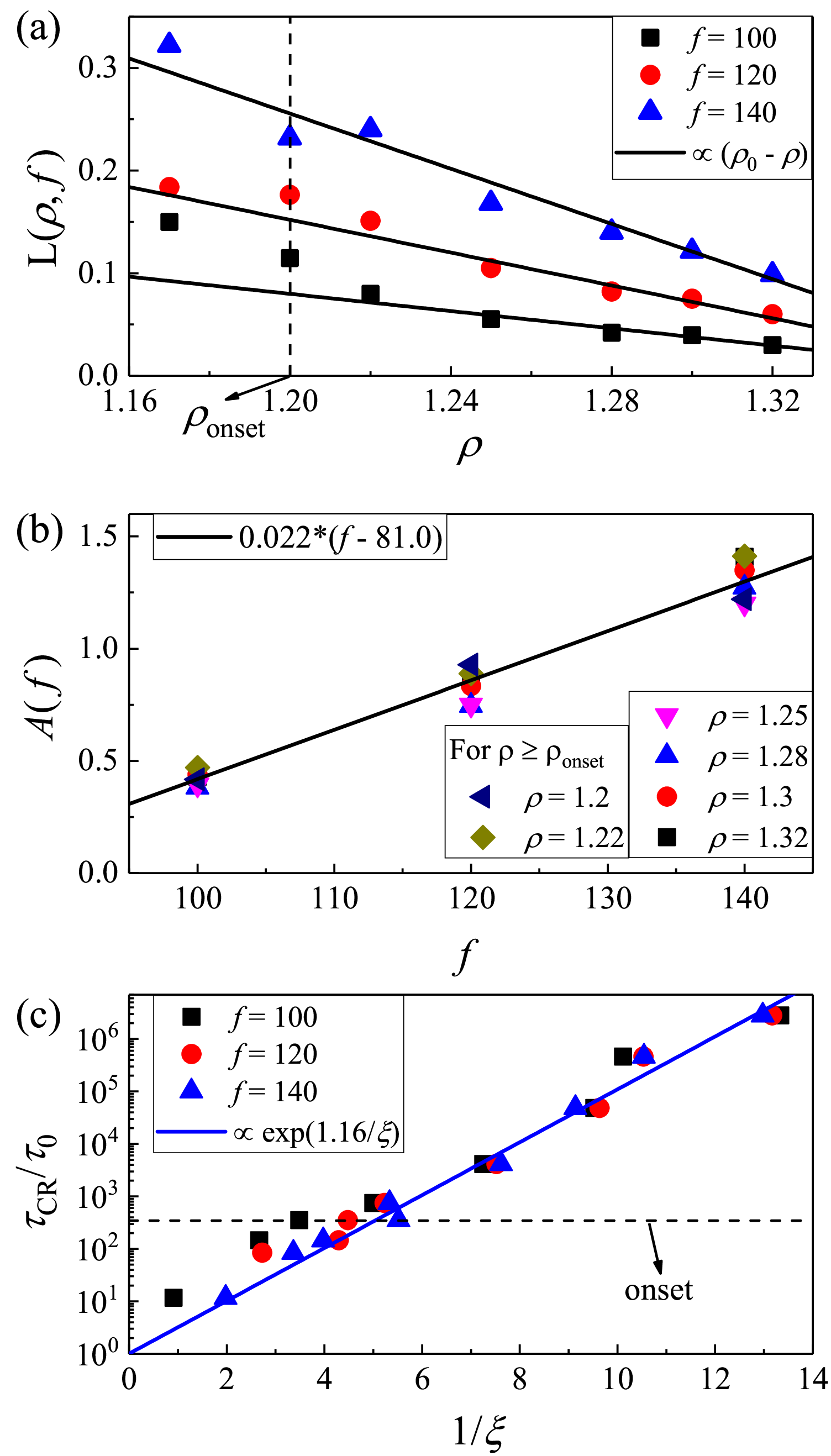}
 \caption{(a) $\rho$ dependence of the asymptotic displacement $L(\rho, f)$ for different values of the applied force $f$. (b) $A(f) = L(T,f)/(T-T_0)$ as a function of f. (c) scaling collapse for $\tau_{\rm CR}/\tau_0$ versus $1/\xi$. The vertical dashed line in (a) and the horizontal one in (c) mark the onset point.
\label{fig:scaling}
}
\end{figure}

For each value of $\rho$, we have performed the investigation described in the main text to extract the typical displacement $L(\rho, f)$ induced by the transient application of a randomly oriented force on a randomly selected particle.
Figure~\ref{fig:scaling}(a) illustrates that $L(\rho, f)= A(f)\xi(\rho)\propto A(f)(\rho_0-\rho)$. 
Here, $A(f)$ depends linearly on $f$ as in Fig.~\ref{fig:scaling}(b). 
Figure~\ref{fig:scaling}(c) demonstrates a data collapse, which supports the scaling relation between relaxation time and plastic length $\tau_{\rm CR}/\tau_0\propto {\rm exp}(1/\xi(\rho))$.
This result demonstrates that the plastic length scale correlates with the relaxation time also for a density-induced slowdown.

\begin{figure}[t!]
 \centering
 \includegraphics[angle=0,width=0.45\textwidth]{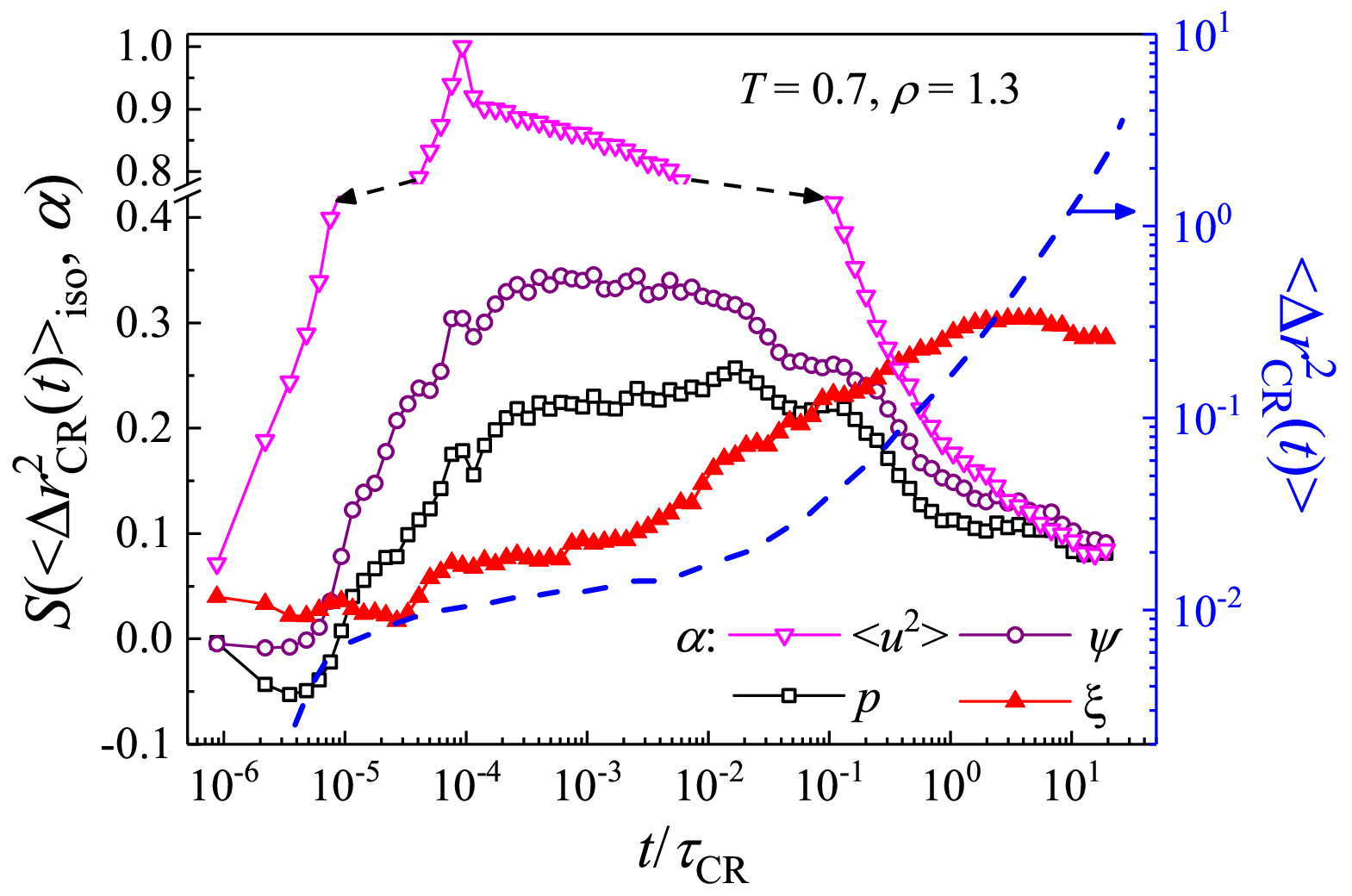}
\caption{The same plot as in Fig. 3 of the main text, here at $T=0.7$ and $\rho=1.3$.
\label{fig:correlation}}
\end{figure}

To assess if these correlations hold at the particle level, we studied the Spearman's rank correlation coefficients between the cage-relative mean square displacement in the iso-configurational ensemble, $\langle\Delta r^2_{i, \rm CR}(t)\rangle_{\rm iso}$ and individual particle properties.
As in Fig. 3 of the main text, we consider the correlations between $\langle\Delta r^2_{i, \rm CR}(t)\rangle_{\rm iso}$ and the Debye-Waller factor $\langle u^2_i\rangle$, the participation ratio $p_i$, and the mean square displacement in the harmonic approximation $\psi_i$.
We perform this analysis at $T=0.7$ and $\rho=1.3$.
Fig.~\ref{fig:correlation} demonstrates that the correlations with $\langle u^2_i\rangle$, $p_i$ and $\psi_i$ are maximal at short times and small at the relaxation time scale. 
Differently, the correlation between $\Delta r^2_{i, {\rm CR}}(t)$ and the plastic length scale $\xi_i$ acquires the maximum at the relaxation time scale.
These findings demonstrate that the plastic length scale is a good predictor of the relaxation dynamics both at the average and at the single-particle level.



~\newpage

\bibliographystyle{apsrev4-1}
\bibliography{SI}